\documentclass[english,pra,aps,showpacs,onecolumn]{revtex4-1}

\def\be{\begin{equation}}
\def\ee{\end{equation}}
\usepackage{graphicx}
\usepackage{bm}
\usepackage[T1]{fontenc}
\usepackage[latin9]{inputenc}
\usepackage{amsmath}
\usepackage{amssymb}
\usepackage{float}
\usepackage{lineno}
\usepackage[bookmarks=true,
   colorlinks=true,
   linkcolor=blue,
   urlcolor=blue,
   citecolor=blue,
   bookmarks=true,
   hyperindex=true
]{hyperref}
\usepackage{natbib}

\begin{document}

\title{Phase-locking matter-wave interferometer of vortex states}

\author{Lingran Kong$^{1,3}$}

\author{Tianyou Gao$^{1}$}
\email{602gty@sina.com}

\author{Longzhi Nie$^{1,3}$}

\author{Dongfang Zhang$^{1}$}

\author{Ruizong Li$^{1,3}$}

\author{Guangwen Han$^{1,3}$}

\author{Mingsheng Zhan$^{1,2}$}

\author{Kaijun Jiang$^{1,2}$}
\email{kjjiang@wipm.ac.cn}

\affiliation{$^{1}$State Key Laboratory of Magnetic Resonance and Atomic and Molecular Physics, Wuhan Institute of Physics and Mathematics, Innovation Academy for Precision Measurement Science and Technology, Chinese Academy of Sciences, Wuhan 430071, China}
\affiliation{$^{2}$Center for Cold Atom Physics, Chinese Academy of Sciences, Wuhan 430071, China}
\affiliation{$^{3}$University of Chinese Academy of Sciences, Beijing 100049, China}

\date{\today}

\begin{abstract}
{\bf
Matter-wave interferometer of ultracold atoms with different linear momenta has been extensively studied in theory and experiment. The vortex matter-wave interferometer with different angular momenta is applicable as a quantum sensor for measuring the rotation, interatomic interaction, geometric phase, etc. Here we report the first experimental realization of a vortex matter-wave interferometer by coherently transferring the optical angular momentum to an ultracold Bose condensate. After producing a lossless interferometer with atoms only populating the two spin states, we demonstrate that the phase difference between the interferences in the two spin states is locked on $\pi$. We also demonstrate the robustness of this out-of-phase relation, which is independent of the angular-momentum difference between the two interfering vortex states, constituent of Raman optical fields and expansion of the condensate. The experimental results agree well with the calculation from the unitary evolution of wave packet in quantum mechanics. This work opens a new way to build a quantum sensor and measure the atomic correlation in quantum gases.}
\end{abstract}

\maketitle

\subsection*{Introduction}
Interference is fundamental to wave dynamics and quantum mechanics. Interferometry represents a unique way to probe the subtle changes of physical parameters by precisely measuring the resultant tiny relative phase shifts. Matter-wave interferometry, especially those realized in ultracold atomic gases, opens a pathway to extract the relative phase between coherent constituents traversing different paths, and holds promise for applications in both practical precision measurement and fundamental quantum research \cite{Pirtchard2009RMPinterferometer}. Like the original optical interferometer where beam splitter (BS) plays a center role, a variety of matter-wave interferometers have emerged employing different mechanisms to realize coherent splitting and recombination of the wavepackets, including double-well potential \cite{Ketterle1997ScienceBECInterference, Schmiedmayer2005NaturePhysicsBECinterference}, Bragg scattering \cite{Phillips2000PRLBraggBECInterferometer}, optical lattice \cite{Oberthaler2010NatureBECInterferometer} and Stern-Gerlach separation \cite{Folman2010NatureCommunicationsBECInterferometer, Folman2015ScienceselfIntefering}, to name a few. As mentioned above, matter-wave interferometer of ultracold atoms with different linear momenta has been extensively studied in theory and experiment. Another type of fundamental matter-wave interferometer would be a vortex matter-wave interferometer with different orbital angular momenta (OAMs). This proposal could be implemented by transferring the OAM of the laser beam to cold atoms through the optical transition, thanks to the recent developments both in vortex light beam carrying definite OAM \cite{Woerdman1992PRALaguerre-GaussianBeam, Allen2003IOP, Yuan2019LightOpticalVortices} and its coherent interaction with cold atoms \cite{Bigelow2009PRLBECVortex, Jiang2019PRLSOAMC, Lin2018PRLSOAMC, Lin2018PRLRotaingBEC}.

Recently, the vortex matter-wave interferometer has been theoretically proposed to be applicable to measure the rotation, magnetic field, interatomic interaction, geometric phase, etc. \cite{Ahufinger2018NJPQuantumSensor, Dowling2012JMOMatterWaveGyroscopy, PRA2016DowlingSagnacInterferlometer, PRResearch2020HornbergerAngularInterference}. Two different vortex states accumulate different phases respect to an external rotation of the system, thus a compact and stable quantum gyroscope without BS and mirror is feasible using this scheme \cite{Dowling2012JMOMatterWaveGyroscopy, PRA2016DowlingSagnacInterferlometer}. The two interfering vortex states can overlap at zero relative velocity, then the long interrogation time can enhance the measurement precision to extract the subtle interatomic interaction \cite{Ahufinger2018NJPQuantumSensor, PRResearch2020HornbergerAngularInterference}. Furthermore, OAM states offer a high dimensional Hilbert space to obtain extra security and dense coding of quantum information \cite{Zeilinger2001NatureOAM, Zeilinger2017PRLOAM}, providing the possibility of using the vortex-state superposition in quantum gases as qubit \cite{Dowling2005PRLvortexQubit, Dowling2005PRAvortexQubit}. People has used the vortex interference patterns in ultracold atoms to measure the winding number of the vortex state \cite{Phillips2006PRLvortex, Bigelow2009PRLBECVortex, Jiang2019PRLSOAMC}. However, the vortex matter-wave interferometer is yet to be explored, where the relative phase between the two interfering vortex states should be quantitatively determined.

Here we report the first experimental realization of a vortex matter-wave interferometer in a two-spin Bose-Einstein condensate. We measure the relative phase between the two interfering vortex states by analyzing the angular interference fringes. A pair of Raman laser beams with an OAM difference as one BS produce the spin-dependent vortex states, and a radio-frequency (RF) pulse as another BS combines two different vortex states into one spin state. After producing two lossless BSs with atoms only populating the two spin states, even the phase of the interference in each spin state fluctuates, the interferences in the two spin states exhibit a constant $\pi$-phase difference. This out-of-phase relation is robust, which is independent of the angular-momentum difference between the two interfering vortex states, constituent of Raman optical fields and expansion of the condensate. The experimental results agree well with the calculation from the unitary evolution of wave packet passing through the lossless BSs in quantum mechanics.

\subsection*{Results}
\begin{figure}[htbp]
\centerline{\includegraphics[width=12cm]{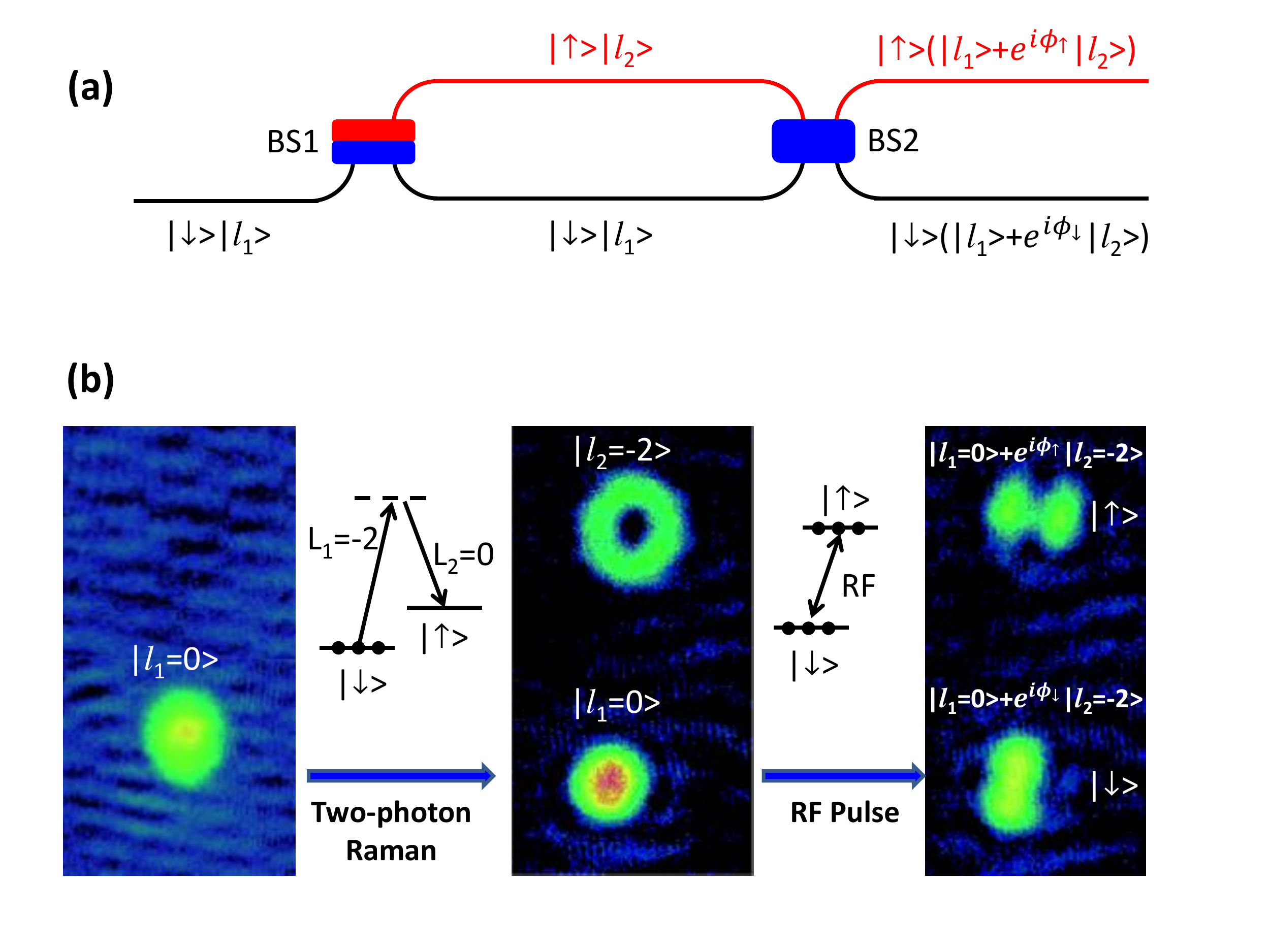}}
\caption{{\bf Schematics of the vortex matter-wave interferometer.} (a) The interferometer is composed of two BSs (BS1 and BS2). The input vortex state $\left|l_{1}\right>$ is in the spin state $\left|\downarrow\right>$. The two-photon Raman pulse, as BS1, contains two optical fields with OAM numbers of $L_{1}$ and $L_{2}$, respectively, producing the vortex state $\left|l_{2}\right>$ in the spin state $\left|\uparrow\right>$. The RF pulse, as BS2, transfers atoms back and forth between the two spin states, generating the interference between the two vortex states $\left|l_{1}\right>$ and $\left|l_{2}\right>$ in each spin state. $\phi_{\uparrow}$ and $\phi_{\downarrow}$ are the phases of the interferences in the two spin states, respectively. (b) An exemplary interferometer with $l_{1} = 0$, $L_{1} = -2$ and $L_{2} = 0$. After BS1, there is a vortex state $\left|l_{1}=0\right>$ in spin state $\left|\downarrow\right>$ and a vortex state $\left|l_{2}=-2\right>$ in spin state $\left|\uparrow\right>$. After BS2, vortex states $\left|l_{1}=0\right>$ and $\left|l_{2}=-2\right>$ interfere in each of the two spin states. }
\label{Fig1}
\end{figure}

\noindent {\bf Scheme of the interferometer.} The vortex matter-wave interferometer is schematically presented in Fig. \ref{Fig1}. The input vortex state with an OAM number of $l_{1}$ is in the spin state $\left|\downarrow\right>$ of the condensate, which can be denoted as $\left|\downarrow\right>\left|l_{1}\right>$. A two-photon Raman pulse composed of a pair of vortex laser beams (with OAM numbers of $L_{1}$ and $L_{2}$, respectively), as BS1, produces the vortex state $\left|l_{2}\right>$ in the spin state $\left|\uparrow\right>$, where $l_{2} = l_{1} + (L_{1}-L_{2})$. After BS1, the quantum states can be written as $\left|\downarrow\right>\left|l_{1}\right> + \left|\uparrow\right>\left|l_{2}\right>$. A RF pulse, as BS2, transfers atoms back and forth between the two spin states, generating the interference between the two vortex states $\left|l_{1}\right>$ and $\left|l_{2}\right>$ in each spin state. After BS2, the quantum state becomes $\left|\uparrow\right>(\left|l_{1}\right> + e^{i\phi_{\uparrow}}\left|l_{2}\right>) + \left|\downarrow\right>(\left|l_{1}\right> + e^{i\phi_{\downarrow}}\left|l_{2}\right>)$. Here we neglect the global phase in each spin state. $\phi_{\uparrow}$ and $\phi_{\downarrow}$ are the relative phases between the two vortex states in spin states $\left|\uparrow\right>$ and $\left|\downarrow\right>$, respectively. Finally a Stern-Gerlach magnetic field pulse makes a projection of the output quantum state onto the two spin states, i.e., the superposition state $\left|l_{1}\right> + e^{i\phi_{\uparrow}}\left|l_{2}\right>$ in $\left|\uparrow\right>$ and $\left|l_{1}\right> + e^{i\phi_{\downarrow}}\left|l_{2}\right>)$ in $\left|\downarrow\right>$. The value of $\phi_{\uparrow, \downarrow}$ can subsequently be extracted from the interference pattern in the two spin states.

Fig. \ref{Fig1} (b) shows an exemplary interferometer with $l_{1} = 0$, $L_{1} = -2$ and $L_{2} = 0$. The vortex states in the two spin states are imaged in three steps of the interferometer, at the input, between the two BSs, and at the output. The spin-resolved atomic images are taken with the help of a Stern-Gerlach magnetic field and a time of flight (TOF) of 20 ms. The petal-like interferences between $\left|l_{1} = 0\right>$ and $\left|l_{2} = -2\right>$ are present in the two output spin states.

Suppose that the two-photon Raman and RF induced transitions are lossless, i.e., atoms only populate the two spin states, then the two BSs can be written as the unitary operators in quantum mechanics \cite{Zeilinger1981AJPGeneralProperties, Dowling2005PRLvortexQubit, Dowling2005PRAvortexQubit}, respectively,

\begin{equation} \label{eq:Raman}
\begin{split}
	 U_{Raman}=
	 \left(\begin{array}{cc}
		 \cos R & -ie^{-i\Delta\phi_{R}}\sin R\left|l_{1}><l_{2}\right|  \\
		 -ie^{i\Delta\phi_{R}}\sin R\left|l_{2}><l_{1}\right| & \cos R
	\end{array} \right),
\end{split}
\end{equation}

\begin{equation} \label{eq:RF}
U_{RF}=\left(\begin{array}{cc}
\cos RF & -ie^{-i\Delta\phi_{RF}}\sin RF \\
-ie^{i\Delta\phi_{RF}}\sin RF & \cos RF
\end{array}\right),
\end{equation}

\noindent where $R=\frac{1}{2}\Omega_{R}T_{R}$ with the Rabi frequency $\Omega_{R}$ and pulse period $T_{R}$ of the optical Raman lights, $RF=\frac{1}{2}\Omega_{RF}T_{RF}$ with the Rabi frequency $\Omega_{RF}$ and pulse period $T_{RF}$ of the RF field. $\Delta\phi_{R}$ and $\Delta\phi_{RF}$ are the obtained phases during the Raman and RF induced transitions, respectively. $U_{Raman}$ transfers the OAM as well as the atomic population between the two spin states, while $U_{RF}$ only transfers the atomic population. As seen in Fig. \ref{Fig1}, the input state is $(\left|\downarrow\right>, \left|\uparrow\right>)^{T} = (u_{10}\left|l_{1}\right>, 0)^{T}$ where $u_{10}$ is the spatial wave function of the initial state. Using Eqs. (\ref{eq:Raman}) and (\ref{eq:RF}), we can obtain the output state of the interferometer,

\begin{equation} \label{eq:interferometer}
\begin{split}
 \left(\begin{array}{c}
u_{1} \\ u_{2}
\end{array}\right) = U_{RF}U_{Raman}\left(\begin{array}{c}
u_{10} \left|l_{1}\right> \\ 0
\end{array}\right)= u_{10}\cdot
 \left(\begin{array}{c}
\cos RF \cos R \left|l_{1}\right> - e^{i\Delta\phi}\sin RF\sin R \left|l_{2}\right> \\
-ie^{i\Delta\phi_{RF}} \left(\sin RF \cos R \left|l_{1}\right> + e^{i\Delta\phi}\cos RF\sin R \left|l_{2} \right> \right)
\end{array}\right),
\end{split}
\end{equation}

\noindent where $\Delta\phi=\left(\Delta\phi_{R}-\Delta\phi_{RF}\right)$. In the following, we set $\Delta\phi$ and $\phi_{\uparrow, \downarrow}$ in the range $\left(-\pi, \pi\right)$. Then $\phi_{\uparrow} = \Delta\phi$ and $\phi_{\downarrow} = \Delta\phi+ (2n + 1)\pi$, where $n = -1, 0$. We obtain the characteristic properties of the lossless vortex matter-wave interferometer: (1) the two BSs can be considered as one lossless BS represented by a unitary operator $U=U_{RF}U_{Raman}$; (2) the interferences in the two spin states have a $\pi$-phase difference,

\begin{equation} \label{eq:phase}
\left|\phi_{\downarrow} - \phi_{\uparrow}\right| = \pi.
\end{equation}

Eq. (\ref{eq:phase}) indicates that, although the relative phase ($\phi_{\uparrow}$ or $\phi_{\downarrow}$) in each spin state can fluctuate due to external coupling with optical Raman and RF pulses, the interferences in the two spin states have a constant $\pi$-phase difference.

The relative phase can be extracted by analyzing the angular interference pattern (see related calculations in Methods). $\theta_{\uparrow}$ and $\theta_{\downarrow}$ are defined as the azimuthal angles of the maximum interference fringes in the two spin states (see Fig. \ref{Fig3}), respectively. With $\Delta l=l_{1}-l_{2}$, the azimuthal angle between the two adjacent maximum interference fringe is equal to $2\pi/|\Delta l|$. Set $\theta_{\uparrow, \downarrow} \in \left(-\pi/|\Delta l|, \pi/|\Delta l|\right)$, then we get the relation of the interference fringes ,

\begin{equation} \label{eq:angle}
\Delta\theta=|\theta_{\downarrow} - \theta_{\uparrow}| = \pi/|\Delta l|.
\end{equation}

\noindent Considering $\phi_{\uparrow}= -\Delta l \theta_{\uparrow}$, $\phi_{\downarrow} = -\Delta l\theta_{\downarrow}$, and $\phi_{\uparrow}-\phi_{\downarrow} = \Delta l(\theta_{\downarrow}-\theta_{\uparrow})$, Eq. (\ref{eq:angle}) exhibits the out-of-phase interferences of Eq. (\ref{eq:phase}). In the following, we will experimentally demonstrate the relation of Eq. (\ref{eq:angle}).

It is noted that, being compared to the linear-momentum state, two vortex states can overlap at zero relative velocity and interfere with a high stability. This unique character provides a convenient way to probe the interference pattern in each spin state.

\begin{figure}[htbp]
\centerline{\includegraphics[width=12cm]{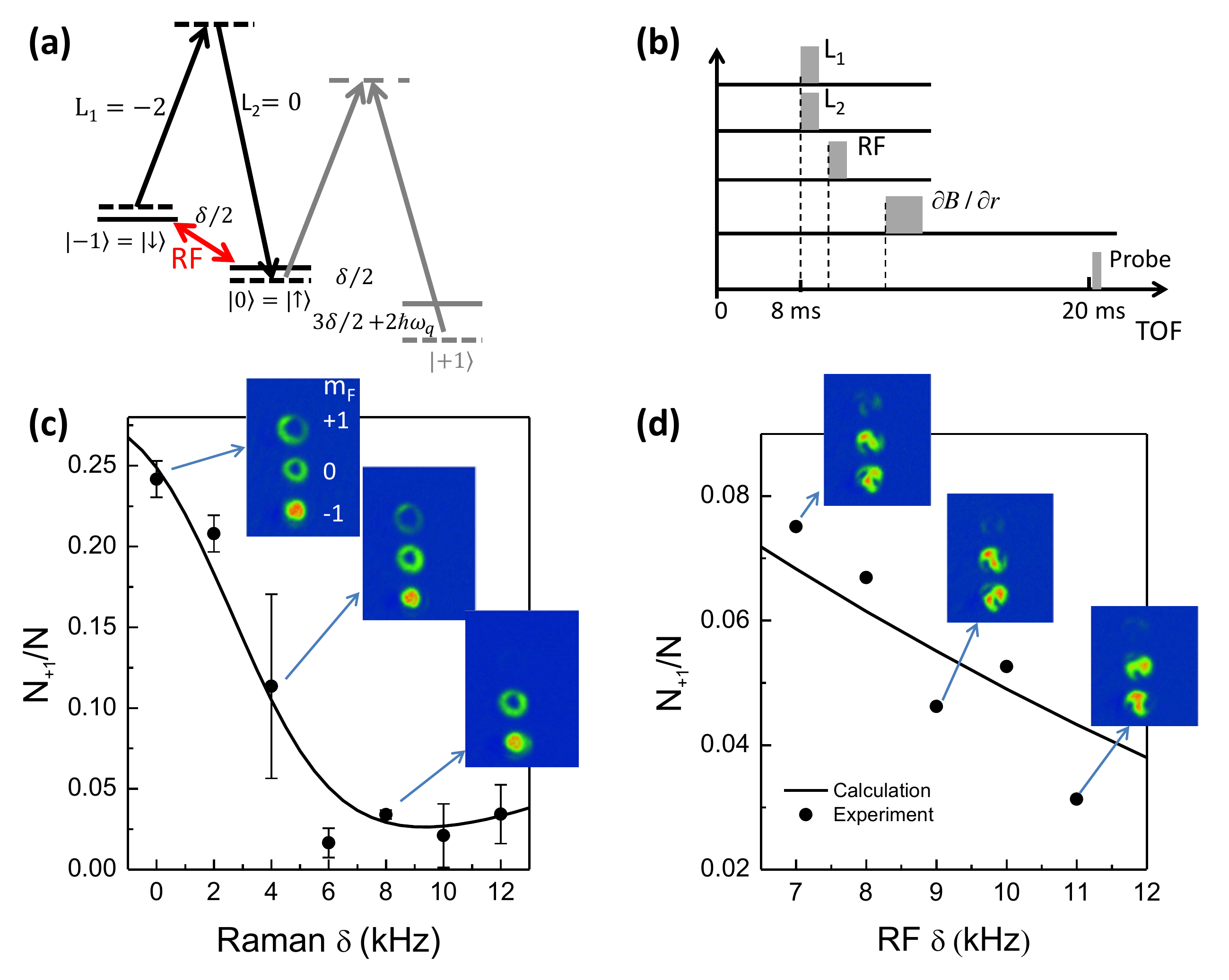}}		
\caption{{\bf Preparation of two lossless BSs (the two-photon Raman pulse and the RF pulse)}. (a) Energy level diagram. Two laser beams couple the two spin states $\left|\uparrow\right>=\left|F=1,m_F=0\right>$ and $\left|\downarrow\right>=\left|F=1,m_F=-1\right>$. $\delta$ is the two-photon detuning. $\omega_q= 2\pi \times 5.52$ kHz is the quadratic Zeeman shift. A RF pulse also couples the two spin states $\left|\downarrow\right>$ and $\left|\uparrow\right>$. (b) The time sequence of the interferometer. After a TOF of 8 ms, the condensate is coupled by a pair of two-photon Raman lights ($L_{1}$ and $L_{2}$) followed with a RF pulse. Then 1 ms later, a gradient magnetic field ($\partial B/\partial r$) is applied to spatially separate different spin states. The spin-resolved atomic density is probed with a TOF of 20 ms. (c) Atom ratio $N_{+1}/N$ versus the detuning $\delta$ of the Raman pulse. $N$ is the total atom number and $N_{+1}$ is the atom number in the spin state $\left|m_{F}=1\right>$. The error bar is the stand deviation of several measurements. Here the RF pulse is absent. The insets show the exemplary atomic images with $\delta=0, 4, 8$ kHz, respectively. The solid curve is the theoretical calculation on three-level atoms coupled with a Raman pulse. (d) Atom ratio $N_{+1}/N$ versus the detuning $\delta$ of the RF pulse. The insets show the exemplary atomic images with $\delta=7, 9, 11$ kHz, respectively. The solid curve is the theoretical calculation on three-level atoms coupled with a RF pulse.}
\label{Fig2}
\end{figure}

\begin{figure*}[htbp]
\centerline{\includegraphics[width=17cm]{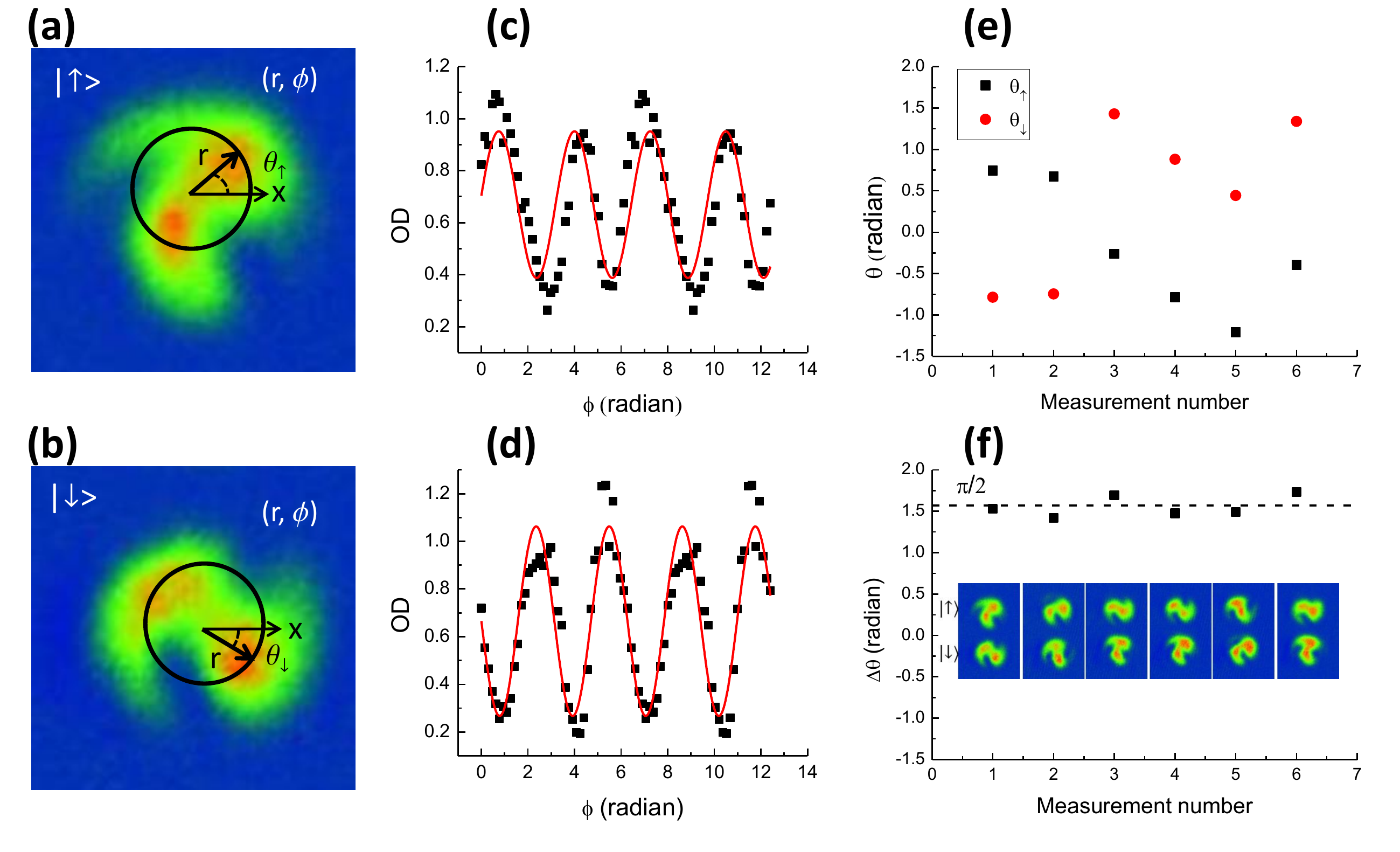}}
\caption{{\bf Out-of-phase interferences in the two spin states for $\Delta l = 2$.} (a) and (b) show exemplary interference patterns between the two vortex states $\left|l_{1}=0\right>$ and $\left|l_{2}=-2\right>$ in the two spin states, respectively. $r$ is the radius, $\phi$ is the azimuthal angle, and $\theta_{\uparrow}$ as well as $\theta_{\downarrow}$ is the azimuthal angle of the maximum interference fringe. (c) and (d) indicate the angular interference fringes along the paths denoted by the black solid circles in (a) and (b), respectively. A cosine wave function (the red solid curve) is used to fit the interference fringe to extract $\theta_{\uparrow, \downarrow}$, where $\theta_{\uparrow}=$ 0.7495, $\theta_{\downarrow}=$ -0.7831 and $\Delta\theta=$ 1.5326. (e) shows the values of $\theta_{\uparrow}$ (black squares) and $\theta_{\downarrow}$ (red circles) for six measurements. $\theta_{\uparrow, \downarrow}$ fluctuates shot to shot. (f) shows a constant difference between the two azimuthal angles, $\Delta\theta = \left|\theta_{\uparrow}-\theta_{\downarrow}\right|\approx\pi/2$. The inset shows the corresponding images of the interference patterns in the two spin states. The first measurement is taken as the example in (a) and (b).}
\label{Fig3}
\end{figure*}

\noindent {\bf Preparation of a lossless interferometer.} We prepare two lossless BSs with atoms only populating the two spin states as shown in Fig. \ref{Fig2}. A $^{87}Rb$ condensate with an atom number of  $N=1.2(1) \times 10^5$ is produced in a nearly spherical optical dipole trap with a trapping frequency $\omega = 2\pi \times 77.5$ Hz \cite{Jiang2019PRLSOAMC, Jiang2019PRBSphericalBEC}. Atoms initially populate the spin state $\left|\downarrow\right>=\left|F=1,m_F=-1\right>$ with zero OAM $l_1=0$. Two laser beams copropagate across the ultracold atoms, transferring the relative OAM of the two laser beams ($L_{1} = -2$ and $L_{2} = 0$) to the condensate in the two-photon Raman induced transition and producing the vortex state $\left|l_{2}=-2\right>$ in the spin state $\left|\uparrow\right>=\left|F=1,m_F=0\right>$ (Fig. \ref{Fig2}(a)). In a previous work \cite{Jiang2019PRLSOAMC}, we have obtained the spin-orbital-angular-momentum coupling (SOAMC) with an adiabatic process in the trap. Here we apply the optical coupling during the expansion of the condensate after a time of 8 ms (see the time sequence in Fig. \ref{Fig2}(b)), enlarging the atomic cloud to increase the coupling strength. The period as well as the power of the Raman and RF pulses are selected to obtain a high interference visibility (see Methods).

$^{87}Rb$ atom has three spin states $\left|m_F=-1, 0, 1\right>$ in the ground state. To produce the lossless Raman and RF induced transitions in which atoms only populate the two spin states $\left|\downarrow\right>$ and $\left|\uparrow\right>$, a bias magnetic field introduces a large quadratic Zeeman shift $\omega_q =2\pi \times 5.52 $kHz and a blue detuning $\delta$ is applied. This will induce a big detuning ($\delta+2\omega_q/2\pi$) of the spin state $\left|m_F= 1\right>$, resulting in a negligible population in it. In Fig. \ref{Fig2}(c), we measure the atom numbers in three spin states versus the two-photon detuning $\delta$ of the Raman pulse. Atom ratio $N_{+1}/N$ decreases when increasing the blue detuning and reaches the minimum at $\delta \approx 9$ kHz, where $N$ is the total atom number and $N_{+1}$ is the atom number in the spin state $\left|m_{F}=1\right>$. In the range $\delta \in$ (8 kHz, 12 kHz), $N_{+1}/N < 0.03$. Considering that the atom number in the spin state $\left|\uparrow\right>$ also decreases with the increased detuning, we generally select a detuning at $\delta \in$ (8 kHz, 10 kHz). In Fig. \ref{Fig2}(d), we measure the atom ratio $N_{+1}/N$ versus the detuning $\delta$ of the RF pulse. $N_{+1}/N$  is smaller than 0.05 in the range $\delta \in$ (9 kHz, 11 kHz). So we can prepare two lossless BSs by choosing the Raman detuning in the range $\delta \in$ (8 kHz, 10 kHz) and the RF detuning in the range $\delta \in$ (9 kHz, 11 kHz).

\noindent {\bf Measurement of the out-of-phase interferences.} In Fig. \ref{Fig3}, we analyze the angular interference fringe between the two vortex states $\left|l_{1}=0\right>$ and $\left|l_{2}=-2\right>$ in the two spin states. As one exemplary measurement, the interference patterns in the two spin states are imaged as shown in Fig. \ref{Fig3}(a) and (b), respectively. In the cylindrical coordinates $(r, \phi, z)$, $r=\sqrt{x^{2}+y^{2}}$ is the radius and $\phi$ is the azimuthal angle in the $x-y$ plane. Cold atoms are confined at $z=0$. $\theta_{\uparrow}$ as well as $\theta_{\downarrow}$ is defined as the azimuthal angle of the maximum interference fringe in each spin state. To extract $\theta_{\uparrow, \downarrow}$, we use a cosine wave function,

\begin{equation} \label{eq:anglefit}
OD=OD_{0}+A\cos\left[|\Delta l|\left(\phi-\theta_{\uparrow, \downarrow}\right)\right] ,
\end{equation}

\noindent to fit the angular interference fringe in Fig. \ref{Fig3}(c) and (d). $\Delta l = l_{1}- l_{2} = 2$, $OD$ is the optical density with a bias $OD_{0}$, and $A$ is the amplitude. It is noted that, to optimize the fitting, two circles with $\phi\in (0, 4\pi)$ are plotted. Because the azimuthal angle between the two adjacent maximum interference fringe is equal to $2\pi/|\Delta l|$, $\theta_{\uparrow, \downarrow}$ is limited in the range $\left(-\pi/|\Delta l|, \pi/|\Delta l|\right)$. Then we get that $\theta_{\uparrow}=$ 0.7495, $\theta_{\downarrow}=$ -0.7831 and $\Delta\theta=$ 1.5326 $\approx\pi/2$. The black solid circle in Fig. \ref{Fig3}(a) or (b) schematically presents the interference path. In order to accurately determine the values of $\theta_{\uparrow, \downarrow}$, we plot 15 interference fringes with a separation of 1 pixel (1 pixel $\approx$ 6.4 $\mu$m) along the radius $r$. Then $\theta_{\uparrow, \downarrow}$ as well as $A$ are plotted versus $r$. We take the values of $\theta_{\uparrow, \downarrow}$ with $A$ being the maximum as the final values. See Supplementary Note 1 for the details.

Despite the fluctuation of the azimuthal angle $\theta_{\uparrow}$ (or $\theta_{ \downarrow}$) shot to shot in each spin state (Fig. \ref{Fig3}(e)), the difference between the two azimuthal angles remains constant, i.e., $\Delta\theta=|\theta_{\uparrow}-\theta_{ \downarrow}|\approx\pi/2$ (Fig. \ref{Fig3}(f)). This is the evidence of the out-of-phase interferences in the two spin states. The Raman and RF pulses will transfer their phases ($\Delta\phi_{R}$ and $\Delta\phi_{RF}$) to the two spin states during their interactions with atoms. The fluctuation of $\Delta\phi_{R}$ and $\Delta\phi_{RF}$ shot to shot results in the variation of the relative phase $\Delta\phi$ (see Eq. (\ref{eq:interferometer})), thus causing the randomness of $\theta_{\uparrow}$ (or $\theta_{ \downarrow}$). We can calculate $\Delta\phi$ as well as $\phi_{\uparrow, \downarrow}$ from each measurement of $\theta_{\uparrow, \downarrow}$ according to the relations $\phi_{\uparrow} = \Delta\phi = -\Delta l\theta_{\uparrow}$ and $\phi_{\downarrow} = -\Delta l\theta_{\downarrow}$ (see Methods). For the measurement in Fig. \ref{Fig3}(a, b), $\phi_{\uparrow}=\Delta\phi=-1.4990$, $\phi_{\downarrow}=1.5662$ and $\phi_{\downarrow} - \phi_{\uparrow} =3.0652$. Then $|\phi_{\downarrow}-\phi_{\uparrow}| \approx\pi$ holds for the six measurements in Fig. \ref{Fig3}(e, f), which also demonstrates the out-of-phase relation. We can calculate the interference patterns with the measured value of $\Delta\phi$. The measured values of $\Delta\phi$ and the calculated interference patterns are shown in Supplementary Note 2.

In Fig. \ref{Fig4}, we demonstrate the out-of-phase interferences for different values of $\Delta l$. We obtain various values of $\Delta l$ by constructing different configurations of the optical Raman pulse, i.e., ($L_{1}=1, L_{2}=-2$) for $\Delta l=-3$, ($L_{1}=0, L_{2}=-2$) for $\Delta l=-2$, ($L_{1}=0, L_{2}=-1$) for $\Delta l=-1$, ($L_{1}=-1, L_{2}=0$) for $\Delta l=1$, ($L_{1}=-2, L_{2}=0$) for $\Delta l=2$, and ($L_{1}=-2, L_{2}=1$) for $\Delta l=3$. The winding number $L_{1, 2}$ of the optical OAM is controlled by a spatial light modulator (SLM). $\theta_{\uparrow, \downarrow}$ as well as $\Delta\theta$ are measured using the same method as in Fig. \ref{Fig3}. A collection of the measurements is shown in Fig. \ref{Fig4}. The theoretical calculation of Eq. (\ref{eq:angle}), $\Delta\theta=\pi/|\Delta l|$, agrees well with the experimental results.

\begin{figure}[htbp]
\centerline{\includegraphics[width=8cm]{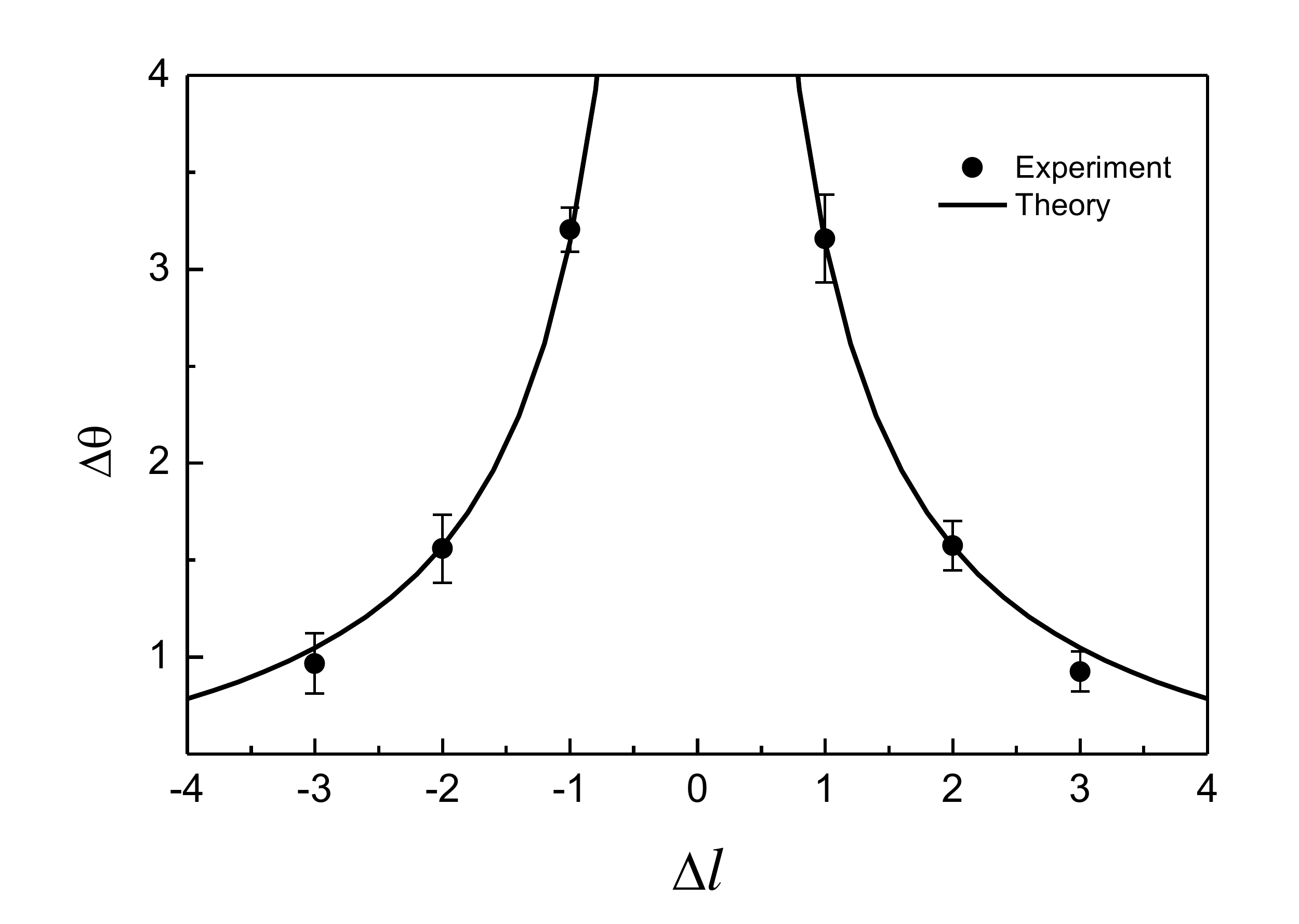}}		
\caption{{\bf Out-of-phase interferences for different $\Delta l$}. $\Delta\theta$ is plotted as a function of $\Delta l$. From left to the right, $\Delta l=-3,-2,-1,1,2,3$. The solid curve denotes the calculation of Eq. (\ref{eq:angle}). The error bar is the standard deviation of six measurements.}
\label{Fig4}
\end{figure}

\begin{figure}[htbp]
\centerline{\includegraphics[width=8cm]{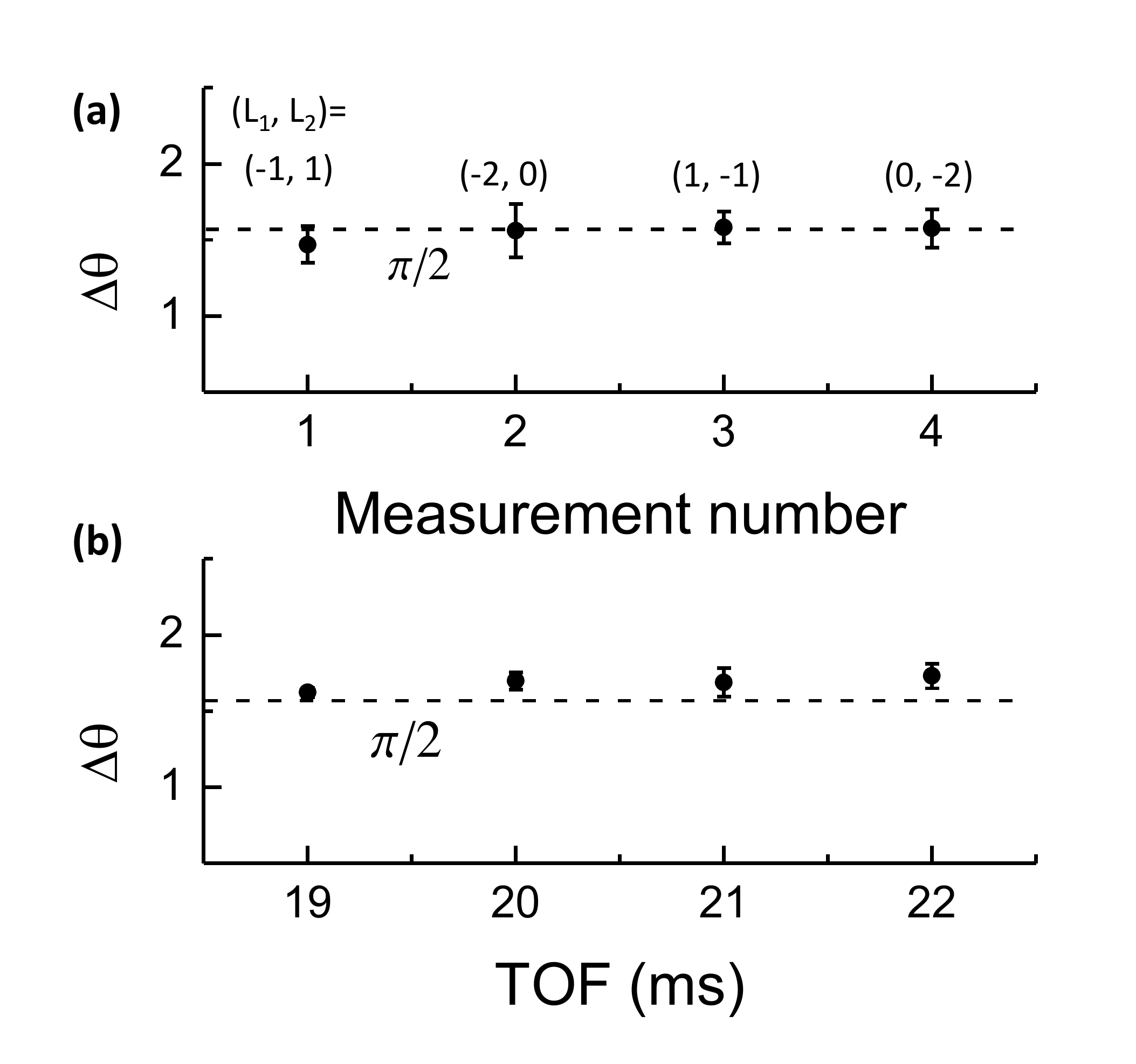}}	
\caption{{\bf Out-of-phase interferences for different constituents of the Raman pulse and during the expansion of the condensate.} (a) shows $\Delta\theta$ for different constituents of the Raman pulse, i.e., ($L_{1}=-1, L_{2}=1$), ($L_{1}=-2, L_{2}=0$), ($L_{1}=1, L_{2}=-1$), and ($L_{1}=0, L_{2}=-2$). $|\Delta l|=2$ and TOF = 20 ms. (b) shows $\Delta\theta$ for different TOFs, i.e., TOF = 19ms, 20ms, 21ms, 22ms. $L_{1}=-2$ and $L_{2}=0$. The dashed line denotes the value of $\pi/2$. The error bar is the standard deviation of six measurements.}
\label{Fig5}
\end{figure}

In Fig. \ref{Fig5}, we demonstrate the out-of-phase interferences for different constituents of the Raman pulse and during the expansion of the condensate. Eq. (\ref{eq:angle}) indicates that $\Delta\theta$ only depends on $|\Delta l|$, but not on the constituent of the Raman pulse. This is analogous to a BS whose property doesn't depend on its composition. In Fig. \ref{Fig5}(a), $|\Delta l|=|L_{2}-L_{1}|=$ 2. Different constituents of the Raman pulse are applied, i.e., ($L_{1}=-1, L_{2}=1$), ($L_{1}=-2, L_{2}=0$), ($L_{1}=1, L_{2}=-1$), and ($L_{1}=0, L_{2}=-2$). $\Delta\theta$ remains constant, $\Delta\theta=\pi/2$. This constant value also holds for different expansion times of the condensate, as indicated in Fig. \ref{Fig5}(b).

\subsection*{Discussion}

In conclusion, we report the first experimental realization of a vortex matter-wave interferometer in ultracold quantum gases. Through producing a lossless interferometer, we demonstrate the out-of-phase relation for the interferences in the two spin states. We further demonstrate the robustness of this out-of-phase relation, which is independent of the angular-momentum difference between the two interfering vortex states, constituent of Raman optical fields and expansion of the condensate. The experimental results agree well with the calculation from the unitary evolution of wave packet in quantum mechanics. In current experiments, despite the fluctuation of the interference phase in each spin state, we show the ability to measure the relative phase between the two vortex states for each measurement. In the future, we will obtain controllable and stable relative phase using optical and RF phase-locking techniques, which is critical to realize a phase-preserving interferometer \cite{Schmiedmayer2005NaturePhysicsBECinterference, Folman2010NatureCommunicationsBECInterferometer}. The vortex matter-wave interferometer is a good candidate to build a quantum sensor for measuring the rotation, magnetic field, interatomic interaction, and geometric phase \cite{Ahufinger2018NJPQuantumSensor, Dowling2012JMOMatterWaveGyroscopy, PRA2016DowlingSagnacInterferlometer, PRResearch2020HornbergerAngularInterference}.

%%% \makeatletter % @ is now a normal "letter" for TeX
%%% \renewcommand{\@biblabel}[1]{[S#1]}
%%% \makeatother % @ is restored as a "non-letter" character for TeX

%%% \nocite{*} %%% show all the referneces including those even not being cited in the paper

\subsection*{Methods}
\noindent \textbf{Experimental setup.} We produce a spherical Rb condensate using the combination of the optical force and gravity as in Ref. \cite{Jiang2019PRLSOAMC, Jiang2019PRBSphericalBEC}. The trapping frequency is $\omega = 2\pi \times 77.5$ Hz. The OAM number is a good quantum number in a system with a rotational symmetry. The atom number is $N=1.2(1) \times 10^5$ and the temperature is $T \approx 50$ nK. The cold atoms initially populate the spin state $\left|\downarrow\right>=\left|F=1,m_F=-1\right>$ with zero OAM $l_1=0$. Two laser beams with different OAMs ($L_1$ and $L_2$) copropagate across the ultracold atoms, transferring the relative OAM of the two laser beams ($\Delta L=L_{1}-L_{2}$) to the condensate in the two-photon Raman induced transition, while suppressing the transfer of the linear momentum. A pair of Helmholtz coils produce a bias magnetic field $B_{0}$, which provides the quantum axis and a large quadratic Zeeman shift $\omega_{q}= 2\pi\times5.52$ kHz of the Rb ground spin states. A pair of anti-Helmholtz coils produce a pulse of a gradient magnetic field $\partial B/\partial r$ to spatially separate different spin states. The probe beam counterpropagates with the Raman beams, detecting the density distribution of the condensate in the plane $r-\phi$.

After a TOF of 8 ms, the condensate is coupled by a pair of two-photon Raman lights ($L_{1}$ and $L_{2}$) followed with a RF pulse. The atom size is about 10 $\mu$m. For the Raman laser beam, the waist is about 70 $\mu$m, the power is about 30 mW. We use the tune-out wavelength $\lambda=$ 790.02 nm of the two Raman beams, in which the ground spin manifold of the Rb atom experiences no scalar ac Stark shift. This can ensure that any vortex structure observed in the condensate is produced due to the OAM transferring, not the trapping effect of the vortex laser beam. The period of the Raman as well as RF pulse is 60 $\mu$s. The Rabi frequency of the optical Raman fields is spatial dependent. The Rabi frequency of the RF field is $2\pi\times1.67$ kHz, which is determined by measuring the Rabi oscillation of the atom numbers in the two spin states. The period as well as the power of the Raman and RF pulses are selected to obtain a high interference visibility. A gradient magnetic field ($\partial B/\partial r$) is applied to spatially separate different spin states. The spin-resolved density is probed with a TOF of 20 ms.

\noindent \textbf{Magnetic field calibration.} The absolute value and stability of the detuning $\delta$ is mainly determined by the bias magnetic field $B_{0}$. We calibrate the magnetic field by adiabatically coupling the ground spin states of $F=1$ with a RF passage. We scan the RF signal from 6.090 MHz to different values in 50 ms and simultaneously record the populations of the three spin states. The Hamitonian of the system dressed by the RF signal is

\begin{equation}
H=
\left (
  \begin{array}{ccc}
 \delta_{RF} & \Omega_{RF}/2  & 0   \\
  \Omega_{RF}/2  & \epsilon & \Omega_{RF}/2    \\
 0 &  \Omega_{RF}/2 & -\delta_{RF}      \\
  \end{array}
\right ), \label{eq:RFcoupling}
\end{equation}

\noindent where $\delta_{RF}/h = \nu_{RF} - \nu_{0}$ is the RF detuning. $h\nu_{0}= (E_{m_{F}=-1}-E_{m_{F}=1})/2$ is the effective resonant position, which is set by the bias magnetic field $B_{0}$. $\Omega_{RF}$ is the coupling strength of the RF signal. $\epsilon=E_{m_{F}=0}-(E_{m_{F}=-1}+E_{m_{F}=1})/2$ is the quadratic Zeeman shift. Then we can numerically calculate relative populations of the three spin states. From the comparison between the experimental measurements and the numerical calculations, we can deduce the bias magnetic field $B_{0}=8.807$ G. By repeating the measurements many times and observing the fluctuation of the resonance position, we can determine the magnetic field stability $\Delta B\approx 1$ mG.

\noindent \textbf{Calculation of the out-of-phase relation.} Here we deduce the out-of-phase relation of Eq. (\ref{eq:angle}). The vortex state is written as $\left|l\right>=e^{-il\theta}$. From Eq. (\ref{eq:interferometer}), the density distributions of the two spin states $\left|\downarrow\right>$ and $\left|\uparrow\right>$ can be written as

\begin{equation} \label{eq:density}
\begin{split}
 |u_{1}|^{2} = u_{10}^{2} [(\cos RF \cos R)^{2} + (\sin RF\sin R)^{2}
 - 2\cos RF \cos R\sin RF\sin R\cos(\Delta\phi+\Delta l\theta)],\\
 |u_{2}|^{2} = u_{10}^{2} [(\sin RF \cos R)^{2} + (\cos RF\sin R)^{2}
 + 2\cos RF \cos R\sin RF\sin R\cos(\Delta\phi+\Delta l\theta)],
\end{split}
\end{equation}

\noindent where $\Delta l=l_{1}-l_{2}$. Then the maximum interference fringes in the two spin states are written as

\begin{equation} \label{eq:maximumfringe}
\begin{split}
& \Delta\phi+\Delta l\theta_{\downarrow} = 2m\pi+\pi,\\
& \Delta\phi+\Delta l\theta_{\uparrow} = 2p\pi,
\end{split}
\end{equation}

\noindent where $m$ as well as $p$ is an integer number. Set $\theta_{\downarrow, \uparrow} \in \left(\frac{-\pi}{\Delta l}, \frac{\pi}{\Delta l}\right)$ and $\Delta\phi \in (-\pi, \pi)$. Then $p=0$ and $m=-1, 0$. We get the relations,

\begin{equation} \label{eq:angledifference}
\begin{split}
\Delta l(\theta_{\downarrow}-\theta_{\uparrow})=\pm\pi,
\end{split}
\end{equation}

\begin{equation} \label{eq:deltaphase}
\begin{split}
\Delta\phi = -\Delta l\theta_{\uparrow}.
\end{split}
\end{equation}

\noindent Eq. (\ref{eq:angledifference}) returns to Eq. (\ref{eq:angle}). From Eq. (\ref{eq:deltaphase}), we can calculate the relative phase $\Delta\phi$ from the measurement of $\theta_{\uparrow}$.

$\phi_{\uparrow}$ and $\phi_{\downarrow}$ are the interferences phases in spin states $\left|\uparrow\right>$ and $\left|\downarrow\right>$, respectively. Set $\phi_{\uparrow, \downarrow} \in (-\pi, \pi)$. From Eq. (\ref{eq:interferometer}), $\phi_{\uparrow} = \Delta\phi$ and $\phi_{\downarrow} = \Delta\phi+ (2n + 1)\pi$, where $n = -1, 0$. In the condition of Eq. (\ref{eq:maximumfringe}),  $\phi_{\uparrow} = -\Delta l\theta_{\uparrow}$ and $\phi_{\downarrow} = -\Delta l\theta_{\downarrow}$. Then we get the relation,

\begin{equation} \label{eq:phaseandangle}
\begin{split}
\phi_{\uparrow}-\phi_{\downarrow} = \Delta l(\theta_{\downarrow}-\theta_{\uparrow}).
\end{split}
\end{equation}

\noindent \textbf{Data availability.} The data that support the findings of this study are available from the corresponding author on request.

\subsection*{Acknowledgments}
We thank Prof. Han Pu for valuable discussions and comments on our manuscript. This work has been supported by the NKRDP (National Key Research and Development Program) under Grant No. 2016YFA0301503, NSFC (Grant No. 11674358, 11904388, 12004398), CAS under Grant No. YJKYYQ20170025, K. C. Wong Education Foundation (Grant No. GJTD-2019-15).

\newpage
\begin{widetext}
\appendix

\setcounter{secnumdepth}{3} %%% Set the number depth as 3, that means the subsubsection is given a digital number. "3" to subsubsection, "2" to subsection

\setcounter{equation}{0}
\setcounter{figure}{0}
\setcounter{table}{0}

\renewcommand\theequation{S\arabic{equation}}
\renewcommand\thefigure{S\arabic{figure}}
\renewcommand\thetable{S\arabic{table}}

\section*{Supplementary Information: Phase-locking matter-wave interferometer of vortex states}

\subsection*{Supplementary Note 1: Measuring the azimuthal angle of the maximum interference fringe} \label{sec:azimuthalAngle}

In Fig. 3 of the main text, we schematically show how to determine the azimuthal angle $\theta_{\uparrow, \downarrow}$ of the maximum interference fringe between $\left|l_{1}=0\right>$ and $\left|l_{2}=-2\right>$ in the two spin states. Here we show the corresponding details in Fig. \ref{FigS1} and Fig. \ref{FigS2}, respectively. First, we take the spin state $\left|\uparrow\right>$ as an example. In order to accurately determine the value of $\theta_{\uparrow}$, we plot 15 interference fringes with a separation of 1 pixel (1 pixel $\approx$ 6.4 $\mu$m) along the radius $\textbf{r}$ (Fig. \ref{FigS1}(a)). We use a cosine wave function,

\begin{equation} \label{eq:15anglefit}
OD=OD_{i0}+A_{i}\cos\left[|\Delta l|\left(\phi_{i}-\theta_{i\uparrow}\right)\right] ,
\end{equation}

\noindent to fit each of the 15 interference fringes, where $i=(3, 4, 5, ....., 17)$ and $r_{i}=(3, 4, 5, ..., 17)$ pixels. $OD$ is the optical density with a bias $OD_{0}$, $A$ is the interference amplitude, and $|\Delta l|=2$. Fig. \ref{FigS1}(b) shows an exemplary interference fringe with $r_{10} = 10$ pixels. It is noted that, to optimize the fitting, two circles with $\phi\in (0, 4\pi)$ are plotted. Because the azimuthal angle between the two adjacent maximum interference fringes is equal to $2\pi/|\Delta l|$, $\theta_{\uparrow}$ is limited in the range $\theta \in \left(-\pi/|\Delta l|, \pi/|\Delta l|\right)$. Then $\theta_{\uparrow}$ as well as $A$ is plotted versus the distance $r$ (Fig. \ref{FigS1}(c) and (d)). We take the value of $\theta_{\uparrow}$ with $A$ being the maximum as the final value. When $r=10$ pixels, $A_{10}=A_{max}=0.28201$ and $\theta_{10 \uparrow}$ = 0.7495. Then we set  $\theta_{\uparrow}=$ 0.7495.

Using the same method, we determine that $\theta_{\downarrow}$ = -0.7831 for the spin state $\left|\downarrow\right>$ (see Fig. \ref{FigS2}). So we get the difference between the azimuthal angles in the two spin states, $\Delta\theta=|\theta_{\uparrow}-\theta_{\downarrow}|=$ 1.5326 $\approx\pi/2$.

\begin{figure}[htbp]
\centerline{\includegraphics[width=8cm]{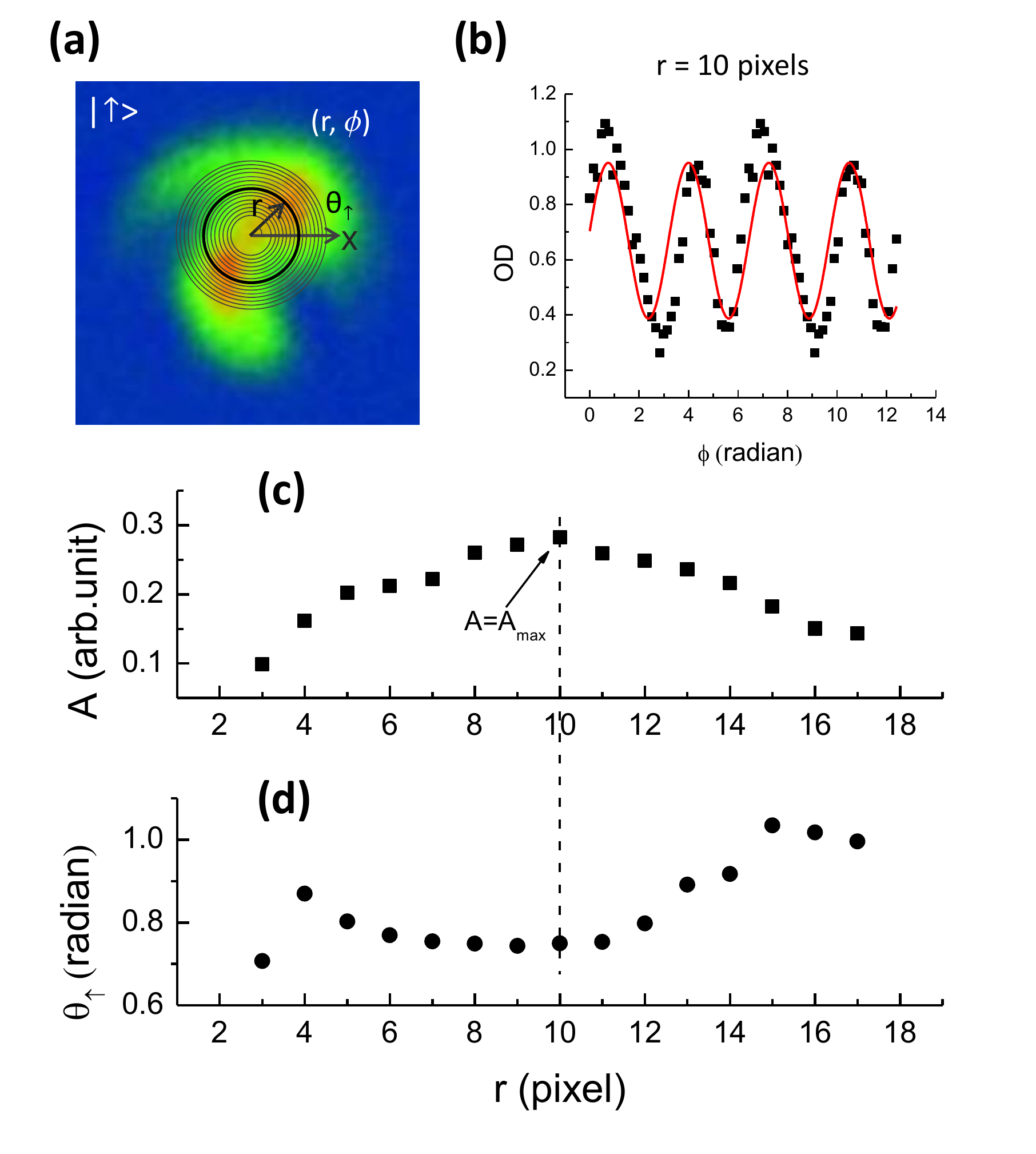}}	
\caption{{\bf Determination of the azimuthal angle $\theta_{\uparrow}$ of the maximum interference fringe in the spin state $\left|\uparrow\right>$.} (a) shows the interference pattern. 15 black circles schematically indicate the interference paths with a separation of 1 pixel (1 pixel $\approx$ 6.4 $\mu$m). (b) shows an exemplary interference fringe with $r=10$ pixels. The red solid curve indicates the numerical fitting with a cosine function of Eq. (\ref{eq:15anglefit}), which gives $\theta_{10\uparrow}$ = 0.7495. (c) and (d) show $\theta_{\uparrow}$ and $A$ as a function of $r$, respectively. As indicated by the vertical dashed line, when $r=10$ pixels, $A_{10}=A_{max}=0.28201$ and $\theta_{10 \uparrow}$ = 0.7495. So we set $\theta_{\uparrow}=$ 0.7495.}
\label{FigS1}
\end{figure}

\begin{figure}[htb]
\includegraphics[width=8cm]{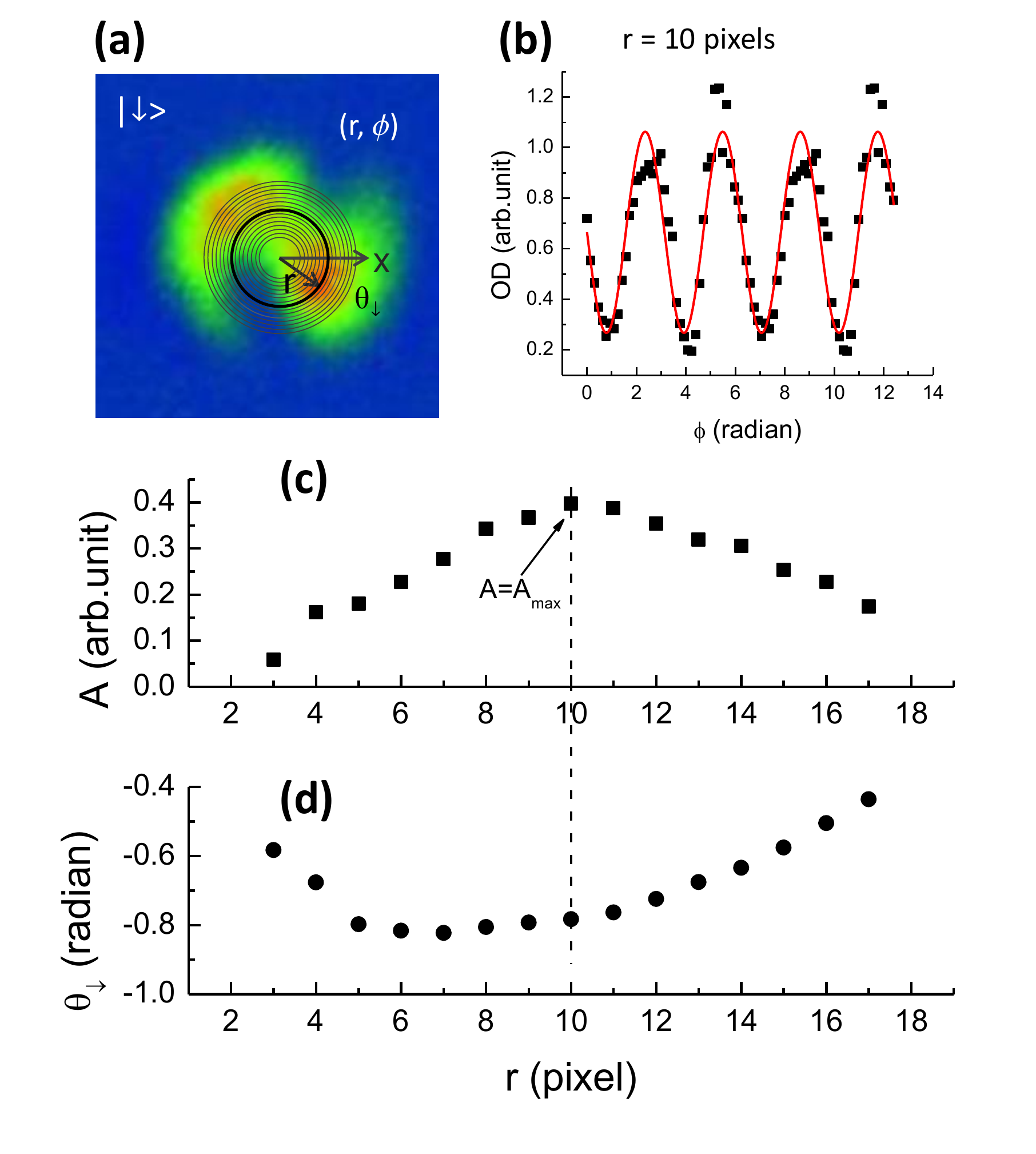}	
\caption{{\bf Determination of the azimuthal angle $\theta_{\downarrow}$ of the maximum interference fringe in the spin state $\left|\downarrow\right>$.} (a) shows the interference pattern. 15 black circles schematically indicate the interference paths with a separation of 1 pixel. (b) shows an exemplary interference fringe with $r=10$ pixels. The red solid curve indicates the numerical fitting with a cosine function of Eq. (\ref{eq:15anglefit}), which gives $\theta_{10 \downarrow}$ = -0.7831. (c) and (d) show $\theta_{\downarrow}$ and $A$ as a function of $r$, respectively. As indicated by the vertical dashed line, when $r=10$ pixels, $A_{10}=A_{max}=0.39747$ and $\theta_{10 \downarrow} = -0.7831$. So we set $\theta_{\downarrow}=$ -0.7831.}
\label{FigS2}
\end{figure}

\subsection*{Supplementary Note 2: Numerical calculation of the interference pattern}

We rewrite Eq. (8) of the main text to calculate the interference patterns in the two spin states,

\begin{equation} \label{eq:densitytwospin}
\begin{split}
& |u_{1}|^{2} = u_{10}^{2} [(\cos RF \cos R)^{2} + (\sin RF\sin R)^{2}  \\
& - 2\cos RF \cos R\sin RF\sin R\cos(\Delta\phi+\Delta l\theta)],\\
& |u_{2}|^{2} = u_{10}^{2} [(\sin RF \cos R)^{2} + (\cos RF\sin R)^{2}  \\
& + 2\cos RF \cos R\sin RF\sin R\cos(\Delta\phi+\Delta l\theta)],
\end{split}
\end{equation}

\noindent where $R=\frac{1}{2}\Omega_{R}T_{R}$ with the Rabi frequency $\Omega_{R}$ and pulse period $T_{R}$ of the optical Raman lights, $RF=\frac{1}{2}\Omega_{RF}T_{RF}$ with the Rabi frequency $\Omega_{RF}$ and pulse period $T_{RF}$ of the RF field. $u_{10}=e^{-r^2/w_{0}^2}$ is the spatial wave function of the initial state, where $w_{0}$ is the size of the condensate. $\Delta\phi$ is the relative phase between the two interfering vortex states. The Rabi Frequency of the optical Raman lights is written as

\begin{equation}
	\Omega_{R} = \Omega_0 (I_{10}I_{20})^{1/2} \left(\frac{r}{w}\right)^{|L_1|+ |L_2|} e^{-2 r^2/w^2},
\end{equation}

\noindent where $L_{1}$ and $L_{2}$ are the winding numbers, $I_{10}$ and $I_{10}$ are the peak intensities, and $w$ is the waist radius.

To calculate the interference pattern, it is required to know the value of $\Delta\phi$. We can extract $\Delta\phi$ from the measurement of $\theta_{\uparrow}$, $\Delta\phi = -\Delta l\theta_{\uparrow}$, according to Eq. (11) of the main text.

In Fig. 3 of the main text, we have measured $\theta_{\uparrow, \downarrow}$ for $\Delta l=2$. Using the same method, we can measure the values of $\theta_{\uparrow, \downarrow}$ for $\Delta l=1, 3$. Then we can  calculate the interference patterns for $\Delta l=2, 1, 3$, which are shown in Fig. \ref{FigS5}. The value of $\Delta\phi$ for each measurement is also given out. All the calculated interference patterns are in good agreements with those measured in the experiment.

\begin{figure*}[htb]
\includegraphics[width=15cm]{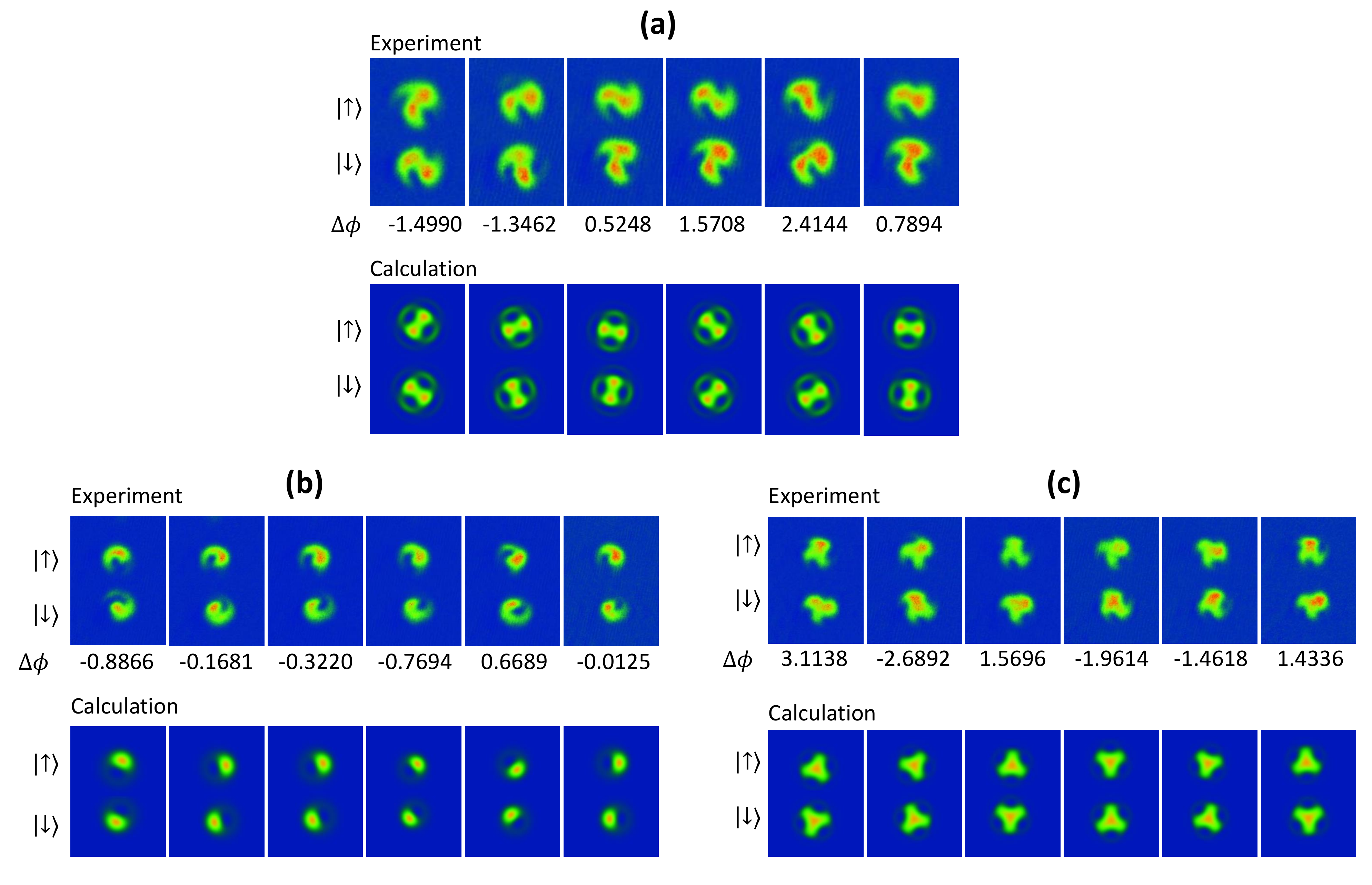}	
\caption{{\bf Numerical calculation of the interference patterns in the two spin states.} The upper panel shows the experimental results of the interference patterns for six measurements. $\Delta\phi$ for each measurement is also given out below. The lower panel shows the theoretical calculations of the interference patterns using Eq. (\ref{eq:densitytwospin}). (a) for $\Delta l=2$. Here $l_{1}=0$, $L_{1} = -2$, $L_{2} = 0$ and $l_{2} = -2$. (b) for $\Delta l=1$. Here $l_{1}=0$, $L_{1} = -1$, $L_{2} = 0$ and $l_{2} = -1$. (c) for $\Delta l=3$. Here $l_{1}=0$, $L_{1} = -2$, $L_{2} = 1$ and $l_{2} = -3$.}
\label{FigS5}
\end{figure*}

\end{widetext}

\end{document}